# Coase's Penguin, or, Linux and the Nature of the Firm

Yochai Benkler[*]

## Abstract

The emergence of GNU/Linux as a viable alternative to the Windows operating system and of the Apache webserver software as the leading web server have focused wide attention on the phenomenon of free or open source software. Most of the attention to the economic aspects of the phenomenon has been focused on the question of incentives—why, it is asked, would anyone invest effort in a productive enterprise in whose fruits they do not claim proprietary rights of exclusion—and has been devoted to studying this phenomenon solely in the context of software development. In this paper I expand consideration of the policy implications of the apparent success of free software in two ways. First, I suggest that the phenomenon has broad implications throughout the information, knowledge, and culture economy, well beyond software development. Second, I suggest reasons to think that peer production may outperform market-based production in some information production activities.

The first part of the paper expands the observation of the phenomenon of peer production to additional layers of the communications process. In particular, I describe instances of peer production of content, relevance, accreditation, and value-added distribution as parallels to peer production of software.

The second part suggests that the primary advantage of peer production is in acquiring and processing information about human capital available to contribute to information production projects, and that in this it is superior to both market-based or hierarchical managerial processes. In addition to the informational advantage, peer production more efficiently assigns human capital to information inputs because it does not rely on controlling bounded sets of either factor. Because of variability of talent and other idiosyncratic characteristics of human capital, there are increasing returns to the scale of the set of agents permitted to work with a set of resources in pursuit of projects, and to the set of resources agents are allowed to work with. The unbounded sets of both human capital and information inputs that are used in peer production capture these economies of scale more effectively than can firms, and to a lesser extent markets, both of which rely on securing access to bounded sets of agents and information inputs to reduce uncertainty about the success of projects.

---

[*] Professor of Law, New York University School of Law. Visiting Professor, Yale Law School. Research for this paper was partly supported by a grant from the Filomen D'Agostino and Max Greenberg Fund at NYU School of Law.

The primary costs of information production and exchange are physical capital costs, communication costs, human capital costs, and information input costs. Information inputs are a pure public good, not an economic good. As rapid advances in computation lower the physical capital cost of information production, and as the cost of communications decline, human capital becomes the salient economic good involved in information production, and the relative advantage of peer production increases in importance.

Peer production is often met by two kinds of objections. The first is the question of incentives. The literature analyzing open source software has largely solved this issue by identifying a series of appropriation mechanisms—from pleasure to reputation and skills development—that serve to justify individual investment in a peer production project. Here I do not add explanations, but rather suggest that the explanations given are examples of a more general claim. That is, the incentives problem is trivial if a sufficient number of individual contributors can be networked, and their various-sized contributions (each determined by the motivation factors driving any given contributor) can be integrated into a finished product. Modularity, granularity of components, and the difficulty/cost of integration become the efficient limit on peer production, not incentives for contribution or the property rights available in the output. This leaves the other traditional wariness of distributed non-market, non-firm based production processes generally captured by the term "tragedy of the commons." I explain how peer production maps on to general understandings of the limitations of commons-based production, and how coordination and integration can be achieved by one of a number of mechanisms, including the reintroduction of limited market-based or hierarchical integration, iterative peer production of integration, and technological clearance. In all events, these options are sufficiently low cost to allow their performance by actors who do not then need to appropriate the whole of the joint product in a way that, if undertaken, would undermine the peer production process *ex ante*.

These conclusions have policy implications for intellectual property law. First, they attenuate the value of strong rights by identifying an important, non proprietary sector that is burdened by, but does not benefit from, strong intellectual property rights. Second, they identify a cost to strong intellectual property rights not previously identified, which is especially salient in a pervasively networked society. This cost is the loss of the information that peer production efforts generate and the loss of potential productive resources—human capital that could be used productively and is not because market-based and hierarchical information production are poor mechanisms for identifying and pooling these resources. This framework suggests avenues for further research.  Peer production offers rich material for empirical studies into collaborative behavior, measurement and comparison of the relative importance of peer production in our information production system, and relative characteristics of information products produced using different organizational approaches.  It also suggests room for theoretical work on organizational forms as information processing devices, under conditions where information processing technology has changed significantly.

### Introduction

Imagine that back in the days when what was good for GM was good for the country an advisory committee of economists had recommended to the President of the United States that the federal government should support the efforts of volunteer communities to design and build their own cars, either for sale or for free distribution to automobile drivers.  The committee members would probably have been locked up in a psychiatric ward—if Senator McCarthy or the House Un-American Activities Committee did not get them first.   Yet, in September of 2000, something like this in fact happened.   The President's Information Technology Advisory Committee recommended that the federal government back open source software as a strategic national choice to sustain the U.S. lead in critical software development.[1]

At the heart of the economic engine of the world's most advanced economies, and in particular that of the United States, we are beginning to take notice of a hardy, persistent, and quite amazing phenomenon—a new model of production has taken root, one that should not be there, at least according to our most widely held beliefs about economic behavior.  It should not, the intuitions of the late 20th century American would say, be the case that thousands of volunteers will come together to collaborate on a complex economic project.  It certainly should not be that these volunteers will beat the largest and best financed business enterprises in the world at their own game.  And yet, this is precisely what is happening in the software world.

The emergence of free software, and the phenomenal success of its flagships—the GNU/Linux operating system, the Apache web server, Perl, sendmail, BIND—and many others,[2] should force us to take a second look at the dominant paradigm we hold about productivity.  That paradigm is that production is organized in only one of two forms—market-based exchanges or firm-based hierarchies.[3]   Both these traditional forms of production depend on clear property rights to control resources and outputs.   This has led information policy in the past decade to focus on strengthening property rights and facilitating contractual exchange—the building blocks of an information economy built on the same model as the economy of coal and steel.   While there are many other critiques of the idea that strong property is appropriate for information and cultural production, the success of open source software offers an additional set of considerations regarding this trend.   This success suggests that in the information economy, the economy that is increasingly occupying center stage in the developed economies, property-based markets and hierarchically managed firms are no longer the only games in town.   Indeed, they might not be the best option.   And the presence of a potentially important mode of production that is harmed, not helped, by property rights should give us more reasons to pause on the way to ever-stronger rights.   But my purpose in this paper is not normative, but positive.   I am trying to outline the parameters of the phenomenon to be studied, and to suggest an initial framework for thinking about it.   More detailed descriptions, and normative conclusions, I leave for future work.

---

[1]  President's Information Technology Advisory Committee, Developing Open Source Software to Advance High End Computing, October, 2000. http://www.ccic.gov/pubs/pitac/pres-oss-11sep00.pdf
[2]  For an excellent history of the free software movement and of the open source development methodology see Glyn Moody, Rebel Code (2001).

Some explanations of free software focus on what is special about software,[4] or about the community ethics of hackers.[5]   Part I of this article will describe a series of similar distributed models of non proprietary production by peers who do not interact either through a firm or through a market.   Indeed, once one begins to look for them, such projects are ubiquitous, robust, and extend to all types of information production and exchange.   They exist in highbrow or lowbrow form.   They cover communicative functions ranging from the authoring of materials to identifying relevance or providing accreditation to materials authored by others, as well as to providing the mundane tasks of the distribution chain, like proofreading.   The phenomenon is, I suggest, generally applicable to information production once you connect everyone to everyone else—an insight encapsulated in Moglen's Metaphorical Corollary to Faraday's Law,[6]—and is explored in Boyle's work on the second enclosure movement[7] and plays a central role in Lessig's argument for embedding the openness of commons in the architecture of the Net.[8]

Part II considers whether from a social welfare perspective peer production is "better" in some measure than market or hierarchical production. The primary answer, following Raymond,[9] given by others who have tried to explain why open source software development is superior, is that it allows development projects to draw on much larger pools of developers—primarily testers who test for bugs, but also programmers who then fix them. Another, more recent explanation, has been that introducing users as part developers and opening the source code permits better information about the eventual end uses desired by consumers, and that such detailed information about specific customization needs is too costly to express in the market, and is therefore lost to commercial manufacturers.[10]

I offer a related but somewhat different explanation, that focuses on the different characteristics of firms, markets, and distributed peer production processes as means of (1) processing information about human capital available to work on projects, and (2) combining human effort and resources to produce outcomes.

My basic claim as to (1) is that different modes of organizing production are better or worse at processing different types of information. Markets and hierarchies are both relatively lossy media when applied to questions of human capital, primarily in terms of creativity and ability, motivation, and focus (efficiency relative to ideal ability at a given time for a given project). This is information that is uniquely in the knowledge of individuals and is immensely difficult to measure and specify in contracts for market clearance or for managerial allocation. A distributed peer production model allows individuals to self-identify for tasks for which they are likely to be the best contributor. This makes peer production particularly well suited to organize activities in which human capital is the dominant input, as long as the coordination problems can be solved and that some mechanism to adjust for individuals' errors regarding their own suitability for a job is implemented.

My basic claim as to (2) is that there are increasing returns to scale when one increases the number of agents who can act on any given resource set, and to the number of resources in a resource set open for any agent to work with.  Markets and firms use property and contract to secure access to bounded sets of agents, resources, and projects, thereby reducing the uncertainty as to the likely effects of decisions by agents outside the sets on the productivity of the agents within the set. The permeability of the boundaries of these sets is limited by the costs of making decisions in a firm about adding or subtracting a marginal resource, agent, or product, and the transaction costs of doing any of these things through the market.  Peer production relies on making an unbounded set of resources available to an unbounded set of agents, who can apply themselves towards an unbounded set of projects and outcomes.   The variability in talent and other idiosyncratic characteristics of individuals suggests that any given resource will be more or less productively used by any given individual, and that the overall productivity of a set of agents and a set of resources will increase more than proportionately when the size of the sets increases towards completely unbounded availability to all agents of all resources for all projects.

---

[7] James Boyle, The Second Enclosure Movement and the Construction of the Public Domain, (paper for the "Conference on the Public Domain," Duke Law School, Durham, North Carolina, November 9-11, 2001).

[8] Lawrence Lessig, The Future of Ideas: The Fate of the Commons in a Connected World (*forthcoming* Random House 2001).

[9] Eric Raymond, The Cathedral and the Bazaar (1998) http://www.tuxedo.org/~esr/writings/cathedral-bazaar/cathedral-bazaar/.

[10] Bessen, *supra*.

The physical capital costs of sophisticated information and cultural production are declining dramatically. Communications systems are making globally distributed talent pools available for any given project while simultaneously reducing the distribution costs of both resources and products.  Under these conditions, two factors of information production gain in salience—the first is a public good, the second a normal economic good.  First are information input costs—the cost of using existing information—like an existing software platform that can be upgraded or customized or an existing database of news items that can be combined into many different commentaries.   The actual marginal cost of these inputs is zero because of the nonrivalry of information, although the price at which they will be available is likely positive, with its precise magnitude a function of the intellectual property system in place and the elasticity of demand for a given input given substitutability with other information inputs.  The second factor, which is the true economic good involved in information production (in the sense that it is both rival and excludable), is human capital.   It is precisely under these conditions—where the other economic goods associated with the production of information—physical capital and communications costs—are low, that the mechanism for allocating human capital becomes central. And it is precisely in the allocation of that factor that the advantage of peer production both as an information process and as an assignment process is salient.

What, then, are the limitations of peer production?  What little economic literature has focused on open source software development has focused on the incentives question.  Why, the question goes, would people invest in producing information if they do not then claim proprietary rights in the product of their labors.[11] In Part II I suggest that—consistent with the claims of those who have attempted to argue for the sustainability of open source development—the incentives problem is a trivial one once a sufficient number of contributors can be involved. The actual limitation on peer production is the capacity of a project to pool a sufficient number of contributors to make the incentives problem trivial. This means that the modularity of an information product and the granularity of its components are what determine the capacity of an information product to be produced on a peer production model. If the components are sufficiently small grained, and the cost of connecting people to projects sufficiently low thanks to efficient cheap network communications, the incentives necessary to bring people to contribute are very small. As a practical matter, given the diversity of motivations and personal valuations on the productive activity itself and on the likelihood of desirable consequences from participation, the incentives problem is trivial.

This leaves the organization problem—how to get all these potential contributors to work on a project together, how to integrate their contributions and control their quality—as the central limiting factor. The organization problems are solved in various forms. These include social organization—like peer review, technical solutions—such as automatic averaging out of outlying contributions, iterative peer production of the integration function itself—such as by developing an open source software solution for integrating contributions, or by a limited reintroduction of market-based and hierarchical mechanisms to produce integration.

## I. Peer Production All Around

While open source software development has captured the attention and devotion of many, it is by no stretch of the imagination the first or most important instance of production by peers who interact and collaborate without being organized on either a market-based or a managerial/hierarchical model. Most important in this regard is the academic enterprise, and in particular science. Thousands of individuals make individual contributions to a body of knowledge, set up internal systems of quality control, and produce the core of our information and knowledge environment. These individuals do not expect to exclude from their product anyone who does not pay for it, and for many of them the opportunity cost of participating in academic research, rather than applying themselves to commercial enterprise, carries a high economic price tag. In other words, individuals produce on a non proprietary basis, and contribute their product to a knowledge "commons" that no one is understood as "owning," and that anyone can, indeed is required by professional norms to, take and extend. We appropriate the value of our contributions using a variety of service-based rather than product-based models (teaching rather than book royalties) and grant funding from government and non-profit sources, as well as, at least as importantly, reputation and similar intangible—but immensely powerful—motivations embodied in prizes, titles etc.   It is easy, though unjustifiable, in the excitement of the moment of transition to an information economy, to forget that information production is one area where we have always had a mixed system of commercial/proprietary and non proprietary peer production—not as a second best or a contingent remainder from the middle ages, but because at some things the non proprietary peer production system of the academic world is simply better.[12]

---

[11] This has been the focus of a number of studies, including those of Raymond, *supra*, Josh Lerner & Jean Tirole, The Simple Economics of Open Source, (2000), http://www.people.hbs.edu/jlerner/simple.pdf; Karim Lakhani & Eric von Hippel, How Open Source Software Works: Free User to User Assistance (2000), http://web.mit.edu/evhippel/www/opensource.PDF, and Moglen *supra* (grudgingly dealing with incentives, a question he deems a reflection of poor understanding of human creative motivation).

In one thing, however, academic peer production and commercial production are similar.   Both are composed of people who are professional information producers.   The individuals involved in production have to keep body and soul together from information production.  However low the academic salary is, it must still be enough to permit one to devote most of one's energies to academic work.  The differences reside in the modes of appropriation and in the modes of organization—in particular how projects are identified and how individual effort is allocated to project. Academics select their own projects, and contribute their work to a common pool that eventually comprises our knowledge of a subject matter, while non-academic producers will often be given their marching orders by managers, who themselves take their focus from market studies, and the product is then sold into the market for which it was produced.

Alongside the professional model, it is also important to recognize that we have always had nonprofessional information and cultural production on a non proprietary model.  Individuals talking to each other are creating information goods, sometimes in the form of what we might call entertainment, and sometimes as a means for news distribution or commentary.  Nonprofessional production has been immensely important in terms of each individual's information environment.  If one considers how much of the universe of communications one receives in a day comes from other individuals in one-to-one or small-scale interactions—such as email, lunch, or hallway conversations—the effect becomes tangible.

---

[12] An early version of this position is Richard R. Nelson, The Simple Economics of Basic Scientific Research, 48 Journal of Political Economy 297-306 (June 1959); more recently one sees the work, for example, of Rebecca S. Eisenberg, Public Research and Private Development: Patents and Technology Transfer In Government-Sponsored Research, 82 Va. L. Rev. 1663, 1715-24 (1996).  For a historical description of the role of market and non-market institutions in science see Paul A. David, From Market Magic to Calypso Science Policy (1997) (Stanford University Center for Economic Policy Research Pub. No. 485).

As computers become cheaper and as network connections become faster, cheaper, and more ubiquitous, we are seeing the phenomenon of nonprofessional peer production of information scale to much larger sizes, performing much more complex tasks than were in the past possible for, at least, nonprofessional production. To make this phenomenon more tangible, I will describe in this part a number of such enterprises, organized so as to demonstrate the feasibility of this approach throughout the information production and exchange chain. While it is possible to break an act of communication into finer-grained subcomponents,[13] largely we see three distinct functions involved in the process. First, there is an initial utterance of a humanly meaningful statement. Writing an article or drawing a picture, whether done by a professional or an amateur, whether high quality or low, is such an action. Then there is a separate function of mapping the initial utterances on a knowledge map. In particular, an utterance must be understood as "relevant" in some sense and "credible." "Relevant" is a subjective question of mapping an utterance on the conceptual map of a given user seeking information for a particular purpose defined by that individual. If I am interested in finding out about the political situation in Macedonia, a news report from Macedonia or Albania is relevant, even if sloppy, while a Disney cartoon is not, even if highly professionally rendered. Credibility is a question of quality by some objective measure that the individual adopts as appropriate for purposes of evaluating a given utterance. Again, the news report may be sloppy and not credible, while the Disney cartoon may be highly accredited *as a cartoon.* The distinction between the two is somewhat artificial, because very often the utility of a piece of information will be as much dependent on its credibility as on its content, and a New York Times story on the Balkans in general will likely be preferable to the excited gossip of a colleague specifically about Macedonia. I will therefore refer to "relevance/accreditation" as a single function for purposes of this discussion, keeping in mind that the two are complementary and not entirely separable functions that an individual requires as part of being able to use utterances that others have uttered in putting together the user's understanding of the world. Finally, there is the function of distribution, or how one takes an utterance produced by one person and distributes it to other people who find it credible and relevant. In the mass media world, these functions were often, though by no means always, integrated. NBC news produced the utterances, by clearing them on the evening news gave them credibility, and distributed them simultaneously. What the Net is permitting is much greater disaggregation of these functions, and so this part will proceed to describe how each component of this information production chain is being produced on a peer-based model on the Net for certain information and cultural goods other than software.

---

[13] See Yochai Benkler, Communications infrastructure regulation and the distribution of control over content, 22 Telecomms Policy 183 (1998).

*1. Content*

NASA Clickworkers is "an experiment to see if public volunteers, each working for a few minutes here and there can do some routine science analysis that would normally be done by a scientist or graduate student working for months on end."[14]  Currently a user can mark craters on maps of Mars, classify craters that have already been marked or search the Mars landscape for "honeycomb" terrain.  The project is "a pilot study with limited funding, run part-time by one software engineer, with occasional input from two scientists."[15]  In its first six months of operation over 85,000 users visited the site, with many contributing to the effort, making over 1.9 million entries (including redundant entries of the same craters, used to average out errors.)  An analysis of the quality of markings showed "that the automatically-computed consensus of a large number of clickworkers is virtually indistinguishable from the inputs of a geologist with years of experience in identifying Mars craters."[16] The tasks performed by clickworkers (like marking craters) are discrete, and  each iteration is easily performed in a matter of minutes. As a result users can choose to work for a minute doing one iteration or for hours by doing many, with an early study of the project suggesting that some  clickworkers indeed work on the project for weeks, but that 37% of the work was done by one-time contributors.[17]

The clickworkers project is a perfect example of how complex and previously highly professional (though perhaps tedious) tasks, that required budgeting the full time salaries of a number of highly trained individuals, can be reorganized so that they can be performed by tens of thousands of volunteers in increments so minute that the tasks can now be performed on a much lower budget.  This low budget is devoted to coordinating the volunteer effort, but the raw human capital needed is contributed for the fun of it.   The professionalism of the original scientists is replaced by a combination of very high modularization of the task, coupled with redundancy and automated averaging out of both errors and purposeful defections (*e.g.,* purposefully erroneous markings).[18]  What the NASA scientists running this experiment had tapped in to was a vast pool of five-minute increments of human intelligence applied with motivation to a task that is unrelated to keeping together the bodies and souls of the agents.

---

[14] http://clickworkers.arc.nasa.gov/top
[15] http://clickworkers.arc.nasa.gov/contact
[16] Clickworkers Results: Crater Marking Activity, July 3, 2001,
http://clickworkers.arc.nasa.gov/documents/crater-marking.pdf
[17] B. Kanefsky, N.G. Barlow, and V.C. Gulick, Can Distributed Volunteers Accomplaish Massive Data Analysis Tasks? http://clickworkers.arc.nasa.gov/documents/abstract.pdf
[18] Clickworkers results, *supra*, para. 2.2.

While clickworkers is a distinct, self-conscious experiment, it suggests characteristics of distributed production that are, in fact, quite widely observable. Consider at the most general level the similarity between this project and the way the World Wide Web can be used to answer a distinct question that anyone can have at a given moment. Individuals put up web sites with all manner of information, in all kinds of quality and focus, for reasons that have nothing to do with external, well-defined economic motives—just like the individuals who identify craters on Mars. A user interested in information need only plug a search request into a search engine like Google, and dozens, or hundreds of websites will appear. Now, there is a question of how to select among them—the question of relevance and accreditation. But that is for the next subpart. For now what is important to recognize is that the web is a global library produced by millions of people. Whenever I sit down to search for information, there is a very high likelihood that someone, somewhere, has produced a usable answer, for whatever reason—pleasure, self-advertising, or fulfilling some other public or private goal as a non-profit or for profit that sustains itself by means other than selling the information it posted. The power of the web comes not from the fact that one particular site has all the great answers. It is not an Encyclopedia Britannica. The power comes from the fact that it allows a user looking for specific information at a given time to collect answers from a sufficiently large number of contributions. The task of sifting and accrediting falls to the user, motivated by the need to find an answer to the question posed. As long as there are tools to lower the cost of that task to a level acceptable to the user, the Web shall have "produced" the information content the user was looking for. These are not trivial considerations (though they are much more trivial today with high speed connections and substantially better search engines than those available a mere two or three years ago). But they are also not intractable. And, as we shall see, some of the solutions can themselves be peer produced.

Another important trend to look at is the emergence and rise of computer games, and in particular multi-player and online games. These fall in the same cultural "time slot" as television shows and movies of the 20th century. The interesting thing about them is that they are structurally different. In a game like Ultima Online, the role of the commercial provider is not to tell a finished, highly polished story to be consumed start to finish by passive consumers. Rather, the role of the game provider is to build tools with which users collaborate to tell a story. There have been observations about this approach for years, regarding MUDs and MOOs. The point here is that there is a discrete element of the "content" that can be produced in a centralized professional manner—the screenwriter here replaces the scientist in the NASA clickworkers example—that can also be organized using the appropriate software platform to allow the story to be written by the many users as they experience it. The individual contributions of the users/co-authors of the storyline are literally done for fun—they are playing a game—but they are spending a real economic good—their attention, on a form of entertainment that displaces what used to be passive reception of a finished, commercially and professionally manufactured good with a platform for active co-production of a storyline. The individual contributions are much more substantial than the time needed to mark craters, but then the contributors are having a whole lot more fun manipulating the intrigues of their imaginary Guild than poring over digitized images of faint craters on Mars.

### 2. Relevance/accreditation

Perhaps, you might say, many distributed individuals can produce content, but it's gobbledygook. Who in their right mind wants to get answers to legal questions from a fifteen-year-old child who learned the answers from watching Court TV?[19] The question, then becomes whether relevance and accreditation of initial utterances of information can itself be produced on a peer production model. At an initial intuitive level the answer is provided by commercial businesses that are breaking off precisely the "accreditation and relevance" piece of their product for peer production. Amazon is a perfect example.

Amazon uses a mix of mechanisms to get in front of their buyers books and other products that the users are likely to buy. It uses a variety of mechanisms to produce relevance and accreditation by harnessing the users themselves. At the simplest level, the recommendation "customers who bought items you recently viewed also bought these items," is a mechanical means of extracting judgments of relevance and accreditation from the collective actions of many individuals who produce the datum of relevance as a by-product of making their own purchasing decisions. At a more self-conscious level (self-conscious, that is, on the part of the user), Amazon allows users to create topical lists, and to track other users as their "friends and favorites," whose decisions they have learned to trust. Amazon also provides users with the ability to rate books they buy, generating a peer-produced rating by averaging the ratings. The point to take home from Amazon is that a corporation that has done immensely well at acquiring and retaining customers harnesses peer production to provide one of its salient values—its ability to allow users to find things they want quickly and efficiently.

Similarly, Google, probably the most efficient general search engine currently in operation, introduced a crucial innovation into ranking results that made it substantially better than any of its competitors. While Google uses a text-based algorithm to retrieve a given universe of web pages initially, its PageRank software employs peer production of ranking in the following way.[20] The engine treats links from other web site pointing to a given website as votes of confidence. Whenever one person who has a page links to another page, that person has stated quite explicitly that the linked page is worth a visit. Google's search engine counts these links as distributed votes of confidence in the quality of that page as among pages that fit the basic search algorithm. Pages that themselves are heavily linked-to count as more important votes of confidence, so if a highly linked-to site links to a given page, that vote counts for more than if an obscure site that no one else thinks is worth visiting links to it. The point here is that what Google did was precisely to harness the distributed judgments of many users, each made and expressed as a byproduct of making one's own site useful, to produce a highly accurate relevance and accreditation algorithm.

---

[19] NY Times Magazine, Sunday, July 15, 2001 cover story.
[20] See description http://www.google.com/technology/index.html.

While Google is an automated mechanism of collecting human judgment as a by product of some other activity (publishing a web page) there are also important examples of distributed projects self-consciously devoted to peer production of relevance. Most prominent among these is the Open Directory Project. The site relies on tens of thousands of volunteer editors to determine which links should be included in the directory. Acceptance as a volunteer requires application, and not all are accepted, and quality relies on a peer review process based substantially on seniority as a volunteer and engagement. The site is hosted and administered by Netscape, which pays for server space and a small number of employees to administer the site and set up the initial guidelines, but licensing is free, and presumably adds value partly to AOL's and Netscape's commercial search engine/portal, and partly through goodwill. The volunteers are not affiliated with Netscape, receive no compensation, and manage the directory out of the joy of doing so, or for other internal or external motivations. Out of these motivations the volunteers spend time on selecting sites for inclusion in the directory (in small increments of perhaps 15 minutes per site reviewed), producing the most comprehensive, highest quality human-edited directory of the Web—competing with, and perhaps overtaking, Yahoo in this category. The result is a project that forms the basis of the directories for many of the leading commercial sites, as well as offering a free and advertising free directory for all to use.

Perhaps the most elaborate mechanism for peer production of relevance and accreditation, at multiple layers, is Slashdot.[21] Billed as "News for Nerds", Slashdot primarily consists of users commenting on initial submissions that cover a variety of technology-related topics. The submissions are typically a link to a proprietary story coupled with some initial commentary from the person who submits the piece. Users follow up the initial submission with comments that often number in the hundreds. The initial submissions themselves, and more importantly the approach to sifting through the comments of users for relevance and accreditation, provide a rich example of how this function can be performed on a distributed, peer production model.

---

[21] http://www.slashdot.org.

First, it is important to understand that the function of posting a story from another site onto Slashdot, the first "utterance" in a chain of comments on Slashdot, is itself an act of relevance production. The person submitting the story is telling the community of Slashdot users "here is a story that people interested in "News for Nerds" should be interested in." This initial submission of a link is itself filtered by "authors" (really editors) who are largely paid employees of Open Source Development Network (OSDN), a corporation that sells advertising on Slashdot. Stories are filtered out if they have technical formatting problems or, in principle, if they are poorly written or outdated. This segment of the service, then, seems mostly traditional—paid employees of the "publisher" decide what stories are, and what are not, interesting and of sufficient quality. The only "peer production" element here is the fact that the initial trolling of the web for interesting stories is itself performed in a distributed fashion. This characterization nonetheless must be tempered, because the filter is relatively coarse, as exemplified by the FAQ response to the question, "how do you verify the accuracy of Slashdot stories?" The answer is, "We don't. You do. If something seems outrageous, we might look for some corroboration, but as a rule, we regard this as the responsibility of the submitter and the audience. This is why it's important to read comments. You might find something that refutes, or supports, the story in the main."[22] In other words, Slashdot very self-consciously is organized as a means of facilitating peer production of accreditation—it is at the comments stage that the story undergoes its most important form of accreditation—peer review *ex post*.

And things do get a lot more interesting as one looks at the comments. Here, what Slashdot allows is the production of commentary on a peer-based model. Users submit comments that are displayed together with the initial submission of a story. Think of the "content" produced in these comments as a cross between academic peer review of journal submissions and a peer-produced substitute for television's "talking heads." It is in the means of accrediting and evaluating these comments that Slashdot's system provides a comprehensive example of peer production of relevance and accreditation.

Slashdot implements an automated system to select moderators from the pool of the users. Moderators are selected according to several criteria; the must be logged in (not anonymous), they must be regular users (selects users who use the site averagely, not one time page loaders or compulsive users), they must have been using the site for a while (defeats people who try to sign up just to moderate), they must be willing, and they must have positive "karma". Karma is a number assigned to a user

---

that primarily reflects whether the user has posted good or bad comments (according to ratings from other moderators). If a user meets these criteria, the program assigns the user moderator status and the user gets five "influence points" to review comments. The moderator rates a comment of his choice using a drop down list with words such as "flamebait" and "informative", etc. A positive word will increase the rating of the comment one point, and a negative word will decrease the rating a point. Each time a moderator rates a comment it costs the moderator one influence point, so the moderator can only rate five comments for each moderating period. The period lasts for three days and if the user does not use the influence points, they expire. The moderation setup is intentionally designed to give many users a small amount of power – thus decreasing the effect of rogue users with an axe to grind or with poor judgment.

The site also implements some automated "troll filters" which prevent users from sabotaging the system. The troll filters prevent users from posting more than once every 60 seconds, prevent identical posts, and will ban a user for 24 hours if the user has been moderated down several times within a short time frame.

Slashdot provides the users with a "threshold" filter that allows each user to block lower quality comments. The scheme uses the numerical rating of the comment (ranging from –1 to 5). Comments start out at 0 for anonymous posters, 1 for registered users and 2 for registered users with good "karma". As a result, if a user sets their filter at 1, the user will not see any comments from anonymous posters unless the comments' ratings were increased by a moderator. A user can set their filter anywhere from –1 (viewing all of the comments), to 5 (where only the posts that have been upgraded by several moderators will show up).

Relevance is also tied into the Slashdot scheme because off topic posts should receive an "off topic" rating by the moderators and sink below the threshold level (assuming the user has the threshold set above the minimum). However, the moderation system is limited to choices that sometimes are not mutually exclusive. For instance, a moderator may have to choose between "funny" (+1) and "off topic" (-1) when a post is both funny and off topic. As a result, an irrelevant post can increase in ranking and rise above the threshold level because it is funny or informative. It is unclear, however, whether this is a limitation on relevance, or indeed mimics our own normal behavior, say in reading a newspaper or browsing a library (where we might let our eyes linger longer on a funny or informative tidbit, even after we've ascertained that it is not exactly relevant to what we were looking for).

Comments are accredited by the rating they receive through moderation. If a user sets a high threshold level, they will only see posts that are considered high quality by the moderators. Users receive accreditation through their karma. If their posts consistently receive high ratings, their karma will increase. At a certain karma level, their comments will start off with a rating of 2 thereby giving them a louder voice in the sense that users with a threshold of 2 will now see their posts immediately. Likewise a user with bad karma from consistently poorly rated comments can lose accreditation by having their posts initially start off at 0 or –1. At the –1 level, the posts will probably not get moderated, effectively removing the opportunity for the "bad" poster to regain any karma.

Together, these mechanisms allow for the distributed production of both relevance and accreditation. Because there are many moderators who can moderate any given comment, and thanks to the mechanisms that explicitly limit the power of any one moderator to over-influence the aggregate judgment, the system evens out differences in evaluation by aggregating judgments. The system then allows users to determine what the level of accreditation pronounced by this aggregate system fits their particular time and needs, by setting their filter to be more or less inclusive, therefore relying to a greater or lesser extent on the judgment of the moderators. By introducing "karma," the system also allows users to build reputation over time, and to gain greater control over the accreditation of their own work relative to the power of the critics. Just, one might say, as very famous authors or playwrights might have over unforgiving critics. The mechanization of means of preventing defection or gaming of the system—applied to both users and moderators—also serve to mediate some of the collective action problems one might expect in a joint effort involving many people.

In addition to the mechanized means of selecting moderators and minimizing their power to skew the aggregate judgment of the accreditation system, Slashdot implements a system of peer-review accreditation for the moderators themselves. Slashdot implements meta-moderation by making any user that has an account from the first 90% of accounts created on the system eligible to moderate the moderations. Each eligible user who opts to perform meta-moderation review is provided with 10 random moderator ratings of comments. The user/meta-moderator then rates the moderator's rating as either unfair, fair, or neither. The meta-moderation process then affects the karma of the original moderator, which, when lowered sufficiently by cumulative judgments of unfair ratings, will remove the moderator from the moderation system.

Users, moderators, and meta-moderators are all volunteers. Using sophisticated software to mediate the multiple judgments of average users regarding the postings of others, the system generates aggregate judgments as to the relevance and quality of user comments. The primary point to take from the Slashdot example is that the same dynamic that we saw used for peer production of content can be implemented to produce relevance and accreditation. Rather than using the full time effort of professional accreditation experts—call them editors or experts—the system is designed to permit the aggregation of many small judgments, each of which entails a trivial effort for the contributor, regarding both relevance and accreditation of the materials sought to be accredited. Another important point is that the software that mediates the communication among the collaborating peers embeds both means to facilitate the participation and to defend the common effort from defection.

### 3. Value-added Distribution

Finally, when we speak of information or cultural goods that exist (content has been produced) and are made usable through some relevance and accreditation mechanisms, there remains the question of "distribution." To some extent this is a non-issue on the Net. Distribution is cheap, all one needs is a server and large pipes connecting one's server to the world, and anyone, anywhere can get the information. I mention it here for two reasons. One, there are a variety of value-adding activities at the distribution stage—like proofreading in print publication—that need to be done at the distribution stage. Again, as long as we are talking about individual web sites, the author who placed the content on the Web will likely, for the same motivations that caused him or her to put the materials together in the first place, seek to ensure these distribution values. Still, we have very good examples of precisely these types of value being produced on a peer production model. Furthermore, as the Net is developing, the largest ISPs are trying to differentiate their services by providing certain distribution–related values. The most obvious examples are caching and mirroring—implementations by the ISP (caching) or a third party like Akamai (mirroring) that insert themselves into the distribution chain in order to make some material more easily accessible than other material. The question is the extent to which peer distribution can provide similar or substitute values.

The most notorious example is Napster. The point here was that the collective availability of tens of millions of hard drives of individual users provided a substantially more efficient distribution system for a much wider variety of songs than the centralized (and hence easier to control) distribution systems preferred by the record industry. The point here is not to sing the praises of the dearly departed (as of this writing) Napster. The point is that, setting aside the issues of content ownership, efficient distribution could be offered by individuals for individuals. Instead of any one corporation putting funds into building a large server and maintaining it, end-users opened part of their hard drives to make content available to others. And while Napster required a central addressing system to connect these hard drives, Gnutella does not. This is not the place to go into the debate over whether Gnutella has its own limitations, be they scalability or free riding. The point is that there are both volunteers and commercial software companies involved in developing software intended to allow users to set up a peer-based distribution system that will be independent of the more commercially-controlled distribution systems, and will be from the edges of the network to its edges, rather than through a controlled middle.

Perhaps the most interesting, discrete and puzzling (for anyone who dislikes proofreading) instantiation of peer-based distribution function is Project Gutenberg and the site set up to support it, Distributed Proofreading. Project Gutenburg[23] entails hundreds of volunteers who scan in and correct books so that they are freely available in digital form. Currently, Project Gutenberg has amassed around 3,500 public domain "etexts" through the efforts of volunteers and makes the collection available to everyone for free. The vast majority of the etexts are offered as public domain materials. The etexts are offered in ASCII format, which is the lowest common denominator and makes it possible to reach the widest audience. The site itself presents the etexts in ASCII format but does not discourage volunteers from offering the etexts in markup languages. It contains a search engine that allows a reader to search for typical fields such as subject, author and title. Distributed Proofreading is a site that supports Project Gutenberg by allowing volunteers to proofread an etext by comparing it to scanned images of the original book. The site is maintained and administered by one person.

Project Gutenberg volunteers can select any book that is in the public domain to transform into an etext. The volunteer submits a copy of the title page of the book to Michael Hart—who founded the project—for copyright research. The volunteer is notified to proceed if the book passes the copyright clearance. The decision on which book to convert to etext is left up to the volunteer, subject to copyright limitations. Typically a volunteer converts a book to ASCII format using OCR (optical character recognition) and proofreads it one time in order to screen it for major errors. The volunteer then passes the ASCII file to a volunteer proofreader. This exchange is orchestrated with very little supervision. The volunteers use a listserv mailing list and a bulletin board to initiate and supervise the exchange. In addition, books are labeled with a version number indicating how many times they have been proofed. The site encourages volunteers to select a book that has a low number and proof it. The Project Gutenberg proofing process is simple and involves looking at the text itself and examining it for errors. The proofreaders (aside from the first pass) are not expected to have access to the book or scanned images, but merely review the etext for self-evident errors.

Distributed Proofreading,[24] a site unaffiliated with the Project Gutenberg, is devoted to proofing Gutenberg Project etexts more efficiently, by distributing the volunteer proofreading function in smaller and more information rich modules. In the

---

[23] http://promo.net/pg/
[24] http://charlz.dynip.com/gutenberg/

Distributed Proofreading process, scanned pages are stored on the site and volunteers are shown a scanned page and a page of the etext simultaneously so that the volunteer can compare the etext to the original page. Because of the modularity, proofreaders can come on the site and proof one or a few pages and submit them. By contrast, on the Project Gutenberg site the entire book is typically exchanged, or at minimum a chapter. In this fashion, Distributed Proofreading clears the proofing of thousands of pages every month.

What is particularly interesting in these sites is that they show that even the most painstaking, and some might say mundane, jobs can be produced on a distributed model. Here the motivation problem may be particularly salient, but it appears that a combination of bibliophilia and community ties suffices (both sites are much smaller and more tightly knit than the Linux development community or the tens of thousands of NASA clickworkers). The point is that individuals can self-identify as having a passion for a particular book, or as having the time and inclination to proofread as part of a broader project they perceive to be in the public good. By connecting a very large number of people to these potential opportunities to produce, Project Gutenberg, just like clickworkers, or Slashdot, or Amazon, can capitalize on an enormous pool of underutilized intelligent human creativity and willingness to engage in intellectual effort.

### 4. Summary

What I hope all these examples provide is a common set of mental pictures of what peer production looks like. In the remainder of the article I will abstract from these stories some general observations about peer production, what makes it work, and what makes it better under certain circumstances than market- or hierarchy-based production. But at this point it is important that the stories have established the plausibility of, or piqued your interest in, the claim that peer production is a phenomenon of much wider application than free software, and that it is something that actually exists, and is not the figment of someone's over-zealous imagination. What remains is the very interesting and difficult task of explaining it in terms that will be comprehensible to those who make economic policy, and that will allow us to begin to think about what, if any, policy implications the emergence of this strange breed in the middle of our information economy has. I will by no stretch of the imagination claim to have completed this task in the following pages. But I hope to identify some basic regularities and organizing conceptions that will be useful to anyone interested in pursuing the answer. Even if you do not buy a single additional word of my initial efforts to theorize the phenomenon, however, seeing these disparate phenomena as in fact instances of a more general phenomenon of uncharacteristic organization of information production should present a rich and fascinating topic of study for organization theorists, anthropologists, institutional economists, and business people interested in understanding production models in a ubiquitously networked environment.

## II. Why Would Peer Production Emerge in a Networked Environment?

### 1. Locating the theoretical space for peer production

There are many places to locate an attempt to provide a theoretical explanation of peer production. One option would be to focus on the literature regarding trust-based modes of organizing production[25] or on literature that focuses on internal motivation and its role in knowledge production.[26] My effort here will be within the more general economics literature that followed Ronald Coase's *The Nature of the Firm* in focusing on the comparative costs of institutional alternatives as an explanation for their emergence and relative prevalence.

At the most general intuitive level, we can begin by looking at Coase's explanation of the firm and Harold Demsetz's explanation of property rights.[27] Coase's basic explanation of why firms emerge—in other words, why clusters of individuals operate under the direction of an entrepreneur, a giver of commands, rather than interacting purely under the guidance of prices—is that using the price system is costly. Where the cost of achieving a certain outcome in the world that requires human behavior through organizational means is lower than the cost of achieving that same result through implementation of the price system, organizations will emerge to attain that result. An organization will cease to grow when (a) another organization can achieve the marginal result that they seek to obtain at lower cost; or (b) the price system can obtain that result at lower cost than can an organization. If the cost of organization increases with size, we have a "natural"—i.e., tractable within welfare economic theory—limit on the size and number of organizations.

Demsetz's basic explanation of why property emerges with regard to resources that previously were managed without property rights—as commons—can be resolved to a very similar rationale. As long as the cost of implementing and enforcing property rights in a given resource is larger than the value of the total increase in the efficiency of the utilization of the resource that would be gained by the introduction of a property regime where none existed before, the resource will operate as a commons. Once the value of the resource increases due to an exogenous circumstance—a technological development or an encounter with another civilization—so that intensification of its utilization through property-based appropriation is worth the cost of implementing property rights, property rights emerge. More generally, this can be stated as: property in a given resource emerges if the social cost of having no property in that resource exceeds the social cost of implementing a property system in it. This restatement can include within it common property regimes, managed commons, and other non-property approaches to managing sustainable commons.[28]

---

[25] See Paul S. Market, Hierarchy, and Trust: The Knowledge Economy and the Future of Capitalism, http://papers.ssrn.com/sol3/papers.cfm?abstract_id=186930
[26] See Margit Osterloh and Bruno S. Frey, Motivation, Knowledge Transfer, and Organizational Form, Motivation, Knowledge Stransfer, and Organizational Form (1999) Institute for Empirical Research in Economics, University of Zurich, Working Paper No. 27.
[27] Harold Demsetz, Toward a Theory of Property Rights, 57 Am. Econ. Rev. 347-357 (1967).

Coase and Demsetz form a four box matrix:

| | Property more valuable than implementation costs[29] | Cost of implementing property higher than opportunity cost of property |
|---|---|---|
| Market exchange of *x* is cheaper than organizing *x* | Pure market | Pure commons? ("market" w/o property => payment by time + effort) |
| Organizing *x* is cheaper than market exchange of *x* | Markets with firms | Common property regimes (if organizing valuable) |

Table 1: Organizational forms as a function of relative social cost of property vs. no-property and firm-based management vs. market

Before going in to *why* peer production may be less costly than property/market based production, or organizational production, it is important to recognize that if we posit the existence of such a third option it is relatively easy to adapt the transactions cost theory of the firm to include it. If we can sustain that under certain circumstances non proprietary or commons-based peer production may be less costly in some dimension than either markets or managerial hierarchies, we could say that when the cost of organizing an activity on a peered basis is lower than the cost of using the market, and the cost of peering is lower than the cost of hierarchical organization, then peer production will emerge.[30]

---

[28] For discussions of commons see, Carol Rose, The Comedy of the Commons: Custom, Commerce, and Inherently Public Property, 53 U. Chi. L. Rev. 711 (1986); Elinor Ostrom, Governing the Commons (1992).

[29] "Valuable" as compared to the option, and opportunity costs, of not having property rights in place.

[30] In the context of land, Ellickson extends Demsetz's analysis in precisely this fashion, suggesting that there may be a variety of reasons supporting group ownership of larger tracks, including the definition of efficient boundaries (efficient for the resource and its use), coping with significant shocks to the resource pool, risk spreading, "the viability of group ownership might be enhanced by the advent of inexpensive video cameras or other technologies for monitoring behavior within a group setting." Robert Ellickson, Property in Land, 102 Yale L.J. 1315, 1330 (1993).

We could tabulate as follows:

|  | No property more costly than implementation costs of property | Property implementation more costly than opportunity cost of no-property |
|---|---|---|
| Market exchange of *x* cheaper than organizing/ peering *x* | Pure market (farmers markets) | Pure commons (ideas & facts; highways?) |
| Organizing *x* cheaper than market exchange or peering of *x* | market with firms | Common property regimes (Swiss pastures) |
| Peering cheaper than both market exchange and organization | Proprietary "open source" efforts (Xerox's Eureka) | Peer production processes[31] (free software; academic science; NASA clickworkers) |

**Table 2: Organizational forms as a function of relative social cost of property vs no-property and firm-based management vs. market vs. peering**

Understanding that in principle the same framework that explains the emergence of property and firms could explain the emergence of peering focuses our effort on trying to understand why it is that peering could under certain circumstances be more cost effective than either markets or hierarchical organizations. To fit the reality of the emergence of peer production in the context of a pervasively networked information economy, that explanation must be (1) in some sense sensitive to changes in the nature of the human and material resources used in production, and (2) affected by the cost and efficiency of communication among human participants in the productive enterprise.

### 2. Peer production of information in a pervasively networked environment

Peer production is emerging as an important mode of information production because of four attributes of the pervasively networked information economy. First, the object of production—information—is quirky, in that (a) it is purely non-rival and (b) its primary non-human input is the same public good as its output—information. Second, the physical capital costs of information production have declined dramatically with the introduction of cheap-processor-based computer networks. Third, the primary human input—creative talent—is highly variable, more so than traditional labor, and the individuals who are the "input" possess better information than anyone else about the variability and suitability of their talents and level of motivation and focus at a given moment to given production tasks. Fourth and finally, communication and information exchange across space and time are much cheaper and more efficient than ever before, which permits the coordination of widely distributed potential sources of creative effort and the aggregation of actual distributed effort into usable end products.

---

[31] "Cost" here would include the negative effects of intellectual property on dissemination and downstream productive use.

The first attribute affects the cost of one major input into production—existing information. It means that the social cost of using existing information as input into new information production is zero. This, on the one hand, lowers the actual social cost of peering by potentially making information input available to human agents without limit, and on the other hand marks a pervasive social cost of market and hierarchy, because of the losses in both static and dynamic efficiency entailed by the property rights in a nonrival public good, usually thought necessary (and hence justifiable) to sustain market and hierarchy-based production. The second attribute similarly lowers the cost of another major capital cost of information production. It was this high cost of physical capital—large circulation automated presses, record and later CD manufacturing and distribution facilities, movie studios and distribution systems, etc.—that formed the basis for the industrial organization of information production that typified commercial cultural production in the 20th century. Together, these two attributes make information production a potentially sustainable low-cost, low returns endeavor for many individuals relying on indirect appropriation.[32] The public goods attribute also limits the applicability of my observations about peer production, so that I make no claim about the applicability of these observations to traditional economic goods. The third characteristic, as I will explain below, is the primary source of efficiency gains from moving from either market or hierarchical organization to peering. Peer production better produces information about available human capital, and increases the size of the sets of agents and resources capable of being combined in projects—where there are increasing returns to scale for these sets. The fourth attribute radically reduces the cost of peering—or coordination of the efforts of many widely distributed participants in a production effort.

What, however, makes contributors to peer production enterprises tick? Why do they contribute? Current explanations of the economics of free software have, I believe, largely resolved the incentives question. The answers fall into two baskets. First, less commonly made, is that people are creative beings. They will play at creation if given an opportunity, and this opportunity to be creative has in fact been seized upon by many.[33] This thesis does not take the question of how to keep body and soul together from this activity as central, but rather explains why people who have a day job will nonetheless devote their time for creative play in this immensely productive manner. This explanation clearly is at work in explaining phenomena like clickworkers or the Slashdot moderators, not only free software. More general explanations are less ambitious for humanity (though possibly less accurate about what motivates people in general) and more ambitious for the stand-alone sustainability of open source software within standard assumptions regarding incentives, because they seek an explanation that relies on the activity itself to keep body and soul together. The primary answers here are that there are a variety of indirect appropriation mechanisms for those who engage in free software development. These range from the amorphous category of reputation gains, through much more mundane benefits such as consulting contracts, customization services, and increases in human capital that are paid for by employers who can use the skills gained from participation in free software development in proprietary projects.[34]

---

[32] "Indirect appropriation" is appropriation of the value of one's effort by means other than reliance on the excludability of the product of the effort. So, someone who is paid as a teacher, but gets the position in reliance on his scholarship, is indirectly appropriating the benefit of his scholarship. An IBM engineer who gains human capital by working on Linux from home in the evening is indirectly appropriating the benefits of her efforts in participating in the production of Linux.

[33] Moglen makes this central to his explanation. Raymond, and Lerner & Tirole also offer hedonic gains as one component of their respective explanations.

The reality of phenomena like academic research, free software, the World Wide Web, NASA's clickworkers or Slashdot supports these explanations with robust, if not quantified, empirical grounding. All one need do is look at the Red Hat millionaires and IBM's billion-dollar commitments to supporting Linux and Apache on the one hand, and the tens of thousands of volunteer clickworkers, thousands of Linux developers, and hundreds of distributed proofreaders, on the other hand, to accept intuitively that some combination of hedonic gain and indirect appropriation can resolve the incentives problem.

But noting that the incentives problem is resolved in this manner does not suggest the limits or characteristics that make some things more amenable to this solution than others. The point to see is that the incentives problem is simple to resolve if the collaboration problem can be solved on a large scale. If a project can draw on the talents of 30,000 or 15,000 individuals instead of a few dozen or a few hundred, then the contribution of each, and hence the personal cost of participation that needs to be covered by indirect benefits, is quite low. In a corollary to "Linus's Law,"[35] one might say

> Given a sufficiently large number of contributions, incentives necessary to bring about contributions are trivial.

Which implies a general statement about where peer production can work:

---

[34] See *supra*, notes 5, 9.

[35] Coined by Eric Raymond to capture one of the attributes of the approach that developed Linux: "Given enough eyeballs, all bugs are shallow."

Peer production is limited not by the total cost or complexity of a project, but by its modularity, the granularity of its components, and the cost of integration.

This is because modularity and granularity determine the minimum individual investment necessary to contribute a component. If this investment is sufficiently low, then incentives for producing a modular component can be of trivial magnitude—like the pure hedonic pleasure of creation. Various actors, with varying levels of motivation, perhaps based on different taste for creation, or on different opportunities for indirect appropriation (like IBM or Red Hat), can collaborate in very different chunks of contributions. The very real question that then remains as an obstacle is the problem of integration of the modules into a finished product. Integration can be solved in some combination of four mechanisms: iterative peer production of the integration function itself, technical solutions embedded in the collaboration platform, norm-based social organization, and limited reintroduction of hierarchy or market to provide the integration function alone.

This general statement leaves us with two questions for the next two sections. The first concerns the relative advantage of peer production over firm-based or market-based production. Even if we were to accept the sustainability of peer production as a matter of "incentives," we would ask whether it is in any dimension "better," so that we can coherently speak of there being cases where using market or hierarchy is more costly—from a general welfare perspective, not from the private perspective of a firm that has outsourced some of its production costs to a bunch of volunteers—than peer production. The second question is how the coordination problems can be solved.

### 3. Markets, hierarchies, and peer production as information processing systems

Peer production has a relative advantage over firm or market-based production along two dimensions, both a function of the highly variable nature of human capital. The first, to which I will devote most of this part, emerges when one treats all approaches to organizing production as mechanisms by which individual agents reduce uncertainty as to the likely value of various courses of action.[36] Differences among these modes in terms of their information processing characteristics could then account for differences in their relative value as mechanisms for organizing production. The second is to see that the particular strategy of reducing uncertainty by attaining secure access to agents and resources through contract and

---

[36] What follows is in some measure a sketchy application of the Herbert Simon's statement, "It is only because individual human beings are limited in knowledge, foresight, skill, and time that organizations are useful instruments for the achievement of human purpose." Simon, Models Of Man 199 (1957).

property—a strategy that firms use extensively and markets to some extent—entails a systematic loss of productivity relative to peer production.

We could reduce the decisions that must be made by productive human beings as follows. Imagine a human agent, $A$, who is a member of a set of human agents $\{A_1, A_2, A_3, \ldots A_n\}$, having to decide to act, where act $a$ is part of the set $\{a_1, a_2, a_3, \ldots a_n\}$. Act $a$ is a combination of two elements: the effort to be exercised, where effort $e$ is part of the set $\{e_1, e_2, e_3, \ldots e_n\}$, representing different levels and focuses of effort possible for $A$, and the resources as to which the effort is exerted, where resource $r$ is part of the set $\{r_1, r_2, r_3, \ldots r_n\}$ available for $A$ to use. Both $e$ and $r$ are sensitive to the costs of collecting information. The components of either set are a function of the set of opportunities to exert effort and the set of resources available to work with that the agent perceives to be open to him. Both sets increase as information collection costs decline, because agents see more of the universe of opportunities actually available to them.

Imagine that $A$ is a rational actor, where the value $V_A$ to $A$ of doing $a$ is the expected value of outcome $O$, where $O$ is part of $\{O_1, O_2, O_3, \ldots O_n\}$, which is the value to $A$ of $O$ obtaining, discounted by the probability that $O$ will obtain if $A$ does $a$. This means that the value to $A$ of doing $a$ increases as the probability that doing $a$ will result in $O$ obtaining increases.

$A_n$ will $a_n$, $(e_n, r_n)$, if the value $V_{An}$ is higher than the value $V_{Am}$ (the value of any other $O_m$ similarly discounted by the probability that any other $a_m$, combining any $(e_m, r_m)$, will lead to $O_m$ obtaining). The value of $O$ and the probability of its obtaining given that any agent $A$ will take any given action $a$ is in some measure dependent on the actions of other agents in $A$'s set. Any such action of another agent is relevant to $A_n$'s decision if it will either interfere with $a_n$ (reduce $V_{An}$ if taken) or complement it (increase $V_{An}$ if taken). The greater the interdependence of the value of an action on the decisions of others regarding their actions—the higher the uncertainty of the value to $A_n$ of any given possible action $a_n$, so long as $A_n$ cannot control the actions of other agents.

Markets and firm-based hierarchies are information processes in the sense that they are means of reducing the uncertainty of agents as to the value of performing one or another productive action—exertion of effort on a set of resources—to a level acceptable to the agent as a level of uncertainty warranting action. Markets will price different levels of effort and resources, so as to signal the relative values of actions in a way that allows individuals to compare actions and calculate the actions of other individuals—whose actions affect the value of the agent's action—faced with similar pricing of alternative courses of action. Firms will reduce uncertainty by specifying to some individuals what actions to take, reducing uncertainty of interdependent action by controlling enough resources and people (by contract and property) to reduce the uncertainty of the outcomes of specified actions to a level acceptable to the managers.

To compare modes of organizing production as information processing systems one might use the term "information opportunity cost." The idea is that different modes of organizing human activity entail different losses of information relative to an ideal state of perfect information. Perfect information is impossible to come by, and different organizational modes have different strategies for overcoming uncertainty, or an absence of perfect information. I use the term "information" here in the technical sense of a reduction in uncertainty, where "perfect information" is the condition where uncertainty regarding an action could not in principle be further reduced.

The different strategies differ from each other in terms of their "lossiness", to use a term from communications. Each loses different amounts or kinds of information in the process of resolving the uncertainty that the lack of perfect information introduces for rational agents deciding what course of action they should follow under given circumstances. This difference among modes of organizing production in terms of the pattern of lossiness is that mode's *information opportunity cost*.

Markets reduce uncertainty regarding allocation decisions by producing a signal that is clear, univocal (i.e., comparable across different uses) as to which use of the relevant factors would be most efficient. To do so, they require a codification of the attributes of different levels of effort, different kinds of resources, and different attributes of outcomes, so that these can all be specified as contract terms to which a price is affixed. An example of this was the introduction of codified standards for commodities as an indispensable element of the emergence of commodities markets.[37]

Since we are concerned with individual agents' decisions, and levels and focuses of effort are a major component of individual action, it is intuitive that specification and pricing of all aspects of individual effort—talent, workload, and focus as they change in small increments over the span of an individual's full day, let alone months—is impossible.[38] What we get instead is codification of effort types—a garbage collector, a law professor—that are priced more-or-less finely. But one need only look at the relative homogeneity of law firm starting salaries as compared to the high variability of individual ability and motivation levels of graduating law students to realize that pricing of individual effort can be quite crude. Moreover, as aspects of performance that are harder fully to specify in advance—like creativity over time given the occurrence over time of opportunities to be creative—become more important, market mechanisms become more lossy.

---

[37] James W. Carey, Communication As Culture (1989).
[38] In the context of the market for labor, this has sometimes been called the multi-task problem (the problem of inability contractually to specify completely all the tasks required and attributes of an employee who will likely need to perform multiple tasks.) See Bengt Holmstrom, The Firm as a Sub-Economy (1999) http://papers.ssrn.com/sol3/papers.cfm?abstract_id=162270.

Markets solve the problem of interdependence in two ways. First, agents can evaluate the risk that others will act in a way that is detrimental, or fail to act in a way that is complementary to, the agent's action, given the relative pricing of the courses of complementary or detrimental action. This risk assessment can then be built into the perceived price of a possible action. Second, agents can maintain property rights in resources and projects,[39] so as to prevent opportunities for negatively correlated action and to provide relatively secure access to resources for complementary action.

Firms or hierarchical organizations resolve uncertainty by instituting an algorithm or process by which information about which actions to follow is ordered, so that some pieces of information lead to a sufficient reduction in uncertainty about the correct course of action to lead to action by factors of production. The mythical entrepreneur (or the historical manager)[40] becomes the sole source of information that is relevant to reducing the uncertainty of the workers in a purely managed firm. In the ideal-type firm, the question of incentives—reducing uncertainty as to which of a set of actions will increase an actor's welfare—is reduced not by reference directly to market signals, but by fixing a salary for a stated behavior (following a manager's orders) and shifting some of the risk of that course of action from employees to employers. Production processes (if I stand here and twist this lever all day, cars will emerge from the other side and I will get a paycheck) are codified as instruction sets. The uncertainty of why to act and what to do (for the human factors of production), or what is to be done with this material factor is reduced by reducing the universe of information relevant to agents' decisions to act. Information that arrives in a particular channel, with a particular level of authorization counts as signal, and all the rest as noise. It remains to the entrepreneur (in the pure model of the firm) to be the interface between the firm and the market, and to translate one set of uncertainty reducing signals—prices—to another set of signals with similar effect—organizational commands.

By controlling a set of resources and commanding a set of agents through property and contract the firm reduces the elements of uncertainty related to the interdependence of the actions of agents. But by doing so it creates a boundary around the set of available agents and set of available resources, and limits the information available about what other agents could have done with these same resources, or what else these agents could have done with these or other resources. This boundary then limits the efficacy of information collection mechanisms—like incentive-based contracts—that firms use to overcome the difficulty of collecting information to which their employees have special access. These mechanisms mean that the employees and resources within the boundary are likely to be better allocated than in firms with no similar mechanism. But firms still lose information about what human agents outside the firm could have done with these resources, or what agents within the firm could have done with resources outside the firm.

---

[39] Maintaining rights in what I call "projects" is, on Kitch's now-classic reading, the primary function of the patent system. See Edmund Kitch, The Nature and Function of the Patent System, 20 J.L. & Econ. 265 (1977). Even if one is critical of Kitch's almost-exclusive focus on this characteristic as the reason for the patent system, recognizing that in some measure patents provide control over projects is all that is necessary here. The derivative use right in copyright plays a similar function to a more limited extent.

[40] Alfred Chandler, The Visible Hand (1977).

The point to see is that like the price system, hierarchical organization is a lossy medium. All the information that could have been relevant to the decision regarding each factor of production, but that was not introduced in a form or at a location that entitled it to "count" towards an agent's decision, given the algorithm used by the organizational structure, is lost. Again, there is an information opportunity cost of using the hierarchical system not only, or even primarily, as compared to the price system. There is an opportunity cost as compared to having perfect information available as to each allocation/action decision of each factor. Much of the knowledge management movement one has seen in business schools since the mid-1990s was concerned with mitigating the lossiness of managerial hierarchy as an information processing mechanism. Mitigating this lossiness is the primary job of CIOs. An example where peer production—though proprietary, not commons-based—was used precisely for this purpose was Xerox's Eureka system for organizing the flow of questions from and answers to field technicians about failures of photocopiers. Creating a database of technician queries, capable of being answered by technicians throughout the Xerox service operations, was a way of capturing some of the lost information by implementing distributed information sharing techniques, rather than hierarchical information flows, to resolve uncertainty about specific actions by technicians. Eureka flipped the traditional conception of knowledge in a machine as codified by engineers and implemented by instruction-following technicians. The knowledge content of the machine was now understood to be something that is incomplete when it leaves the design board, and is completed over the life of the machines by technicians who share questions and solutions on a peer-review, volunteer model over a proprietary communications platform.

Recognizing the lossiness of markets and managerial hierarchies suggests a working hypothesis about why peer production has succeeded in gaining ground. Peer production may have sufficiently lower information opportunity costs as compared to markets and hierarchies, that the gains it offers as an information collection and processing mechanism outweigh the costs of overcoming the coordination problems among widely distributed peers whose behavior is not mediated by price or controlled by contract.

One of the things that makes the "effort" component of an agent's decision to act difficult to specify sufficiently to get very information-rich pricing or managerial decision-making is that individuals are immensely complex and diverse in terms of their raw abilities, focus (capacity to translate raw ability into effective action) and motivation (willingness to exert effort and focus at a given time for a given project). All these also change not only from individual to individual, but also within one individual over time and specific external conditions. Codifying the desired attributes becomes increasingly difficult and lossy. So, for example, if the task required is very simple and routine—standing at a particular spot for eight hours and turning a knob—then the set of individual attributes necessary to specify is minimal, and an individual willing and able to perform it can be identified very efficiently through pricing or hierarchy. The passion of that same individual for Mars, and her motivation and capacity to look at Mars craters and identify them for fifteen minutes a week will likely go unappreciated and untapped in a hierarchy-based organization, and quite possibly in a price system as well. It is knowledge she may only come to possess after she has encountered the NASA clickworkers project, to which she then begins to contribute in her evenings.

The point is more general: human intellectual effort is highly variable and individuated. It is very difficult to standardize and specify in contracts—necessary for either market-cleared or hierarchically organized production. As human intellectual effort increases in importance as an input into a given production process, an organization model that does not require a standard contractual specification of the effort required to participate in a collective effort and allows individuals to self-identify for tasks will be better at clearing information about who should be doing what than a system that does require such specification.

So, where the physical capital costs of information production are low and information itself is freely accessible (that is, where intellectual property law has not raised the price of using inputs above its marginal cost of zero), and where the cost of communication is low, individual agents can relatively cheaply scour the universe of resources to look for opportunities to invest their creative capabilities. (Their incentive to do so is any one of the indirect appropriation mechanisms ranging from pleasure to reputation etc.) Individuals can then identify both opportunities for creative utilization of the existing resources in ways not previously done,[41] and opportunities to use their own talents, availability, focus, and motivation to perform a productive act. Systems that lower the cost of coordination among peers, clearing information about what needs to or can be done, who is doing it, etc., will then enable these individuals to contribute variously-sized contributions, each of which is qualitatively better than could have been matched by markets or hierarchies. To succeed, such systems also require a mechanism for smoothing out incorrect self-assessments—as peer review does, or as, in the case of NASA clickworkers, simple redundancy and statistical averaging does. The prevalence of misperceptions as to one's ability, and the cost of eliminating such errors, will be part of the transaction costs associated with this form of organization that are parallel to quality control problems faced by firms and markets.[42] This problem is less important where the advantage of peer production is in acquiring fine-grained information about motivation and availability of individuals with widely available intellectual capabilities—like the ability to evaluate the quality of someone else's comment on Slashdot. It is likely more important where what is necessary is a particular skill set that may not be widespread—like fixing a bug in a program.

---

[41] This is a point Bessen makes about complex software, see Bessen, *supra*, as well as a characteristic of the motivation Raymond describes as having an itch to scratch.

The point here is qualitative. It is not only, or even primarily that more people can participate in production (as is sometimes said of open source development). It is that the widely distributed model of information production will better identify who is the best person to produce a specific modular component, *all abilities and availability to work on the specific module within a specific time frame* considered.

With enough uncertainty as to the value of various productive activities and enough variability in the quality of both information inputs and human creative talent vis-à-vis any set of production opportunities, coordination and continuous communications among the pool of potential producers and consumers can generate better information about the most valuable productive actions, and the best human inputs available to engage in these most valuable actions at a given time.

Peer production has an additional advantage relative to markets and firms, that is a function of the use that firms, and to a lesser extent market actors, make of boundaries to reduce uncertainty as to the availability of resources and agents. This advantage is cumulative to the information-processing characteristics of peer production, and relates to the size of the pools of individuals, resources, and projects engaged in information production. The same problem that creates the informational advantage of peer production—the high variability of human capital—also suggests that there are increasing returns to scale in increasing the pool of individuals and resources potentially available for use in any given production project. As explained, market- and particularly firm-based production management rely on property and contract to secure access to bounded sets of agents and resources in the pursuit of specified projects. The permeability of the boundaries of these sets is limited by the costs of making decisions in a firm about adding or subtracting a marginal resource, agent, or product, or the transaction costs of doing any of these things through the market. Peer production relies on making an unbounded set of resources available to an unbounded set of agents, who can apply themselves towards an unbounded set of projects and outcomes. The variability in talent and other idiosyncratic characteristics of individuals suggests that any given resource will be more or less productively used by any given individual, and that the overall productivity of a set of agents and a set of resources will increase more than proportionately when the size of the sets increases towards completely unbounded availability of all agents to all resources for all projects. The point is that even if in principle we have information as to who was the best person for a job given any particular set of resources and projects, the transaction or organizational costs involved in bringing that agent to bear on the project may be too great relative to the efficiency gain over use of the resource by the next-best available agent who is within the boundary.

---

[42] I owe thanks to Doug Lichtman for pointing out the importance of individual self-misperception as a particular potential failure of the information processing characteristics of peer production, that must be solved by these processes if they are to succeed.

Assume that the productivity ($P$) of a set of agents/resources is a function of the Agents ($A$) available to invest effort on resources ($r$). $P_{A1}$ (Productivity of agent $A_1$) is a function of set of resources $A_1$ can work on, $r_1$, the level of effort of A, $e_1$, and A's talent, ($t_1$) where talent describes relative capabilities, associations, and idiosyncratic insights and educational mixes of an individual. $P_A$ increases as $r$ and $e$ increase, at a magnitude that is a function of $t$. $t$ is a personal characteristic of individuals that is independent of the set of resources open for $A$ to work on, but will make a particular $A$ more or less likely to be effective with a given set of resources for the achievement of a set of outcomes. The existence of $t$ means that there are increasing returns to making a larger set of resources available to a larger set of agents, because the larger the number of agents with access to a larger number of resources the higher the probability that the agents will include someone, $A_x$, with relatively high value of $t$, who will be more productive with $r_x$ given some $e_x$ than other agents. If $A_1$ who works for $F_1$ has a higher $t$ value as regards using $r_2$, than $A_2$ who works for $F_2$, but $r_2$ is owned by $F_2$, $r_2$ will nonetheless be used by $A_2$ so long as the value of $A_2$ working on $r_2$ has a value of no less than the value of $A_1$ working on $r_2$ minus the transaction costs involved in assigning $A_1$ to $r_2$. This potential efficiency loss would be eliminated if $A_1$ were in the set of agents who had baseline access to work on the resource set that includes $r_2$. So, if firms $F_1$ and $F_2$ each has a set of agents and resources, $\{A_{F1}, r_{F1}\}$ $\{A_{F2}, r_{F2}\}$, then $P_{F1} + P_{F2} < P_{F1+2}$. Furthermore, a more diverse set of talents looking at a set of resources may reveal available projects that would not be apparent when one only considers the set of resources as usable by a bounded set of agents. Finally, none of this takes into consideration collaboration, and the possible ways in which collaborating individuals can make each other creative in different ways than they otherwise might have been. Once one takes into consideration these diverse effects on the increased possibilities for relationships among individuals and between individuals and resources, it becomes more likely that there are substantial increasing returns to increases in the number of agents and resources involved in a production process.[43]

If this is true, then in principle a state in which all agents can act on all resources effectively will be substantially more productive in creating information goods than a world in which firms divide the universe of agents and resources into bounded sets. As peer production relies heavily on opening up access to resources, for a relatively unbounded set of agents, freeing them to define and pursue an unbounded set of projects that are the best outcome of combining a particular individual or set of individuals with a particular set of resources, this largely open set of agents is likely to be more productive than the same set could have been if divided into bounded sets in firms. If the modularity of a product is insufficient to permit the aggregation of low levels of effort, and more directed incentives are necessary to induce effort, this effect might be muted relative to the importance of the application of direct incentives-based effort. In other words, the ability to link compensation to effort may be more important than the efficiency loss created by introducing a firm or market capable of inducing that incentives-based effort. But if effort/incentives can be at least partially solved by decentralization, the substantial increases in productivity born of the availability of a larger set of resources to a larger set of agents with widely variable talent endowments could be enough to make even an imperfectly motivated peer production process more productive than firms that more directly motivate effort but segment agents and resources into smaller bounded sets.

### 4. Integration: Problem and opportunity

This leaves us with a consideration of the last factor limiting peer production—the possibility and cost of integration of distributed efforts into a common product. It is here that the term "commons" in describing the phenomenon as "commons-based peer production," gets its bite. Its role is to denote the centrality of the absence of property rights as a central organizing feature of this new mode of production, and to evoke the potential pitfalls of such an absence for decentralized production efforts.

What kind of commons is it, then, that peer production of information relies upon? Commons are most importantly defined by two parameters.[44] The first parameter is whether use of the resource is common to everyone in the world or to a well-defined subset. The term "commons" is better reserved for the former, while the latter is better identified as a common property regime (CPR)[45] or limited common property regime.[46] The second parameter is whether use of the resource by whoever the set of people whose use is privileged is regulated or not. Here one can more generally state, following Rose, that resources in general can be subject to regimes ranging from total (and inefficiently delineated) exclusion—the phenomenon Heller has called the anticommons[47]—through efficiently-delineated property and otherwise regulated access, to completely open, unregulated access.[48] The infamous "tragedy of the commons" is best reserved to refer only to the case of unregulated access commons, whether true commons or CPRs. Regulated commons need not be tragic at all, and indeed have been sustained and shown to be efficient in many cases.[49] The main difference here is that CPRs are usually easier to monitor and regulate—using both formal law and social norms[50]—than true commons, hence the latter may more often slip into the open access category even when they are formally regulated.

---

[43] I am not sure there is room to formalize the relationship here on the style of Metcalfe's Law or Reed's Law, see David P. Reed, That Sneaky Exponential—Beyond Metcalfe's Law to the Power of Community Building, http://www.reed.com/Papers/GFN/reedslaw.html. From a policy perspective, there is no need to do so at this early stage of studying the phenomenon. It is sufficient for our purposes here to see that the collaboration effects and insights due to exposure to additional resources mean that the returns to scale are more than proportional.

Ostrom also identified that one or both of two economic functions will be central to the potential failure or success of any given commons-based production system. The first is the question of provisioning, the second of allocation. This may seem trivial, but it is important to keep the two problems separate, because if a particular resource is easily renewable if allocated properly then institutions designed to assure provisioning would be irrelevant. Fishing and whaling are examples. In some cases, provisioning may be the primary issue. Ostrom describes various water districts that operate as common property regimes that illustrate well the differences between situations where allocation of a relatively stable (but scarce) water flow exists, on one hand, and where provisioning of a dam is the difficult task, after which water is relatively abundant.[51]  Obviously, some commons will require both.

---

[44] The most extensive consideration of commons and the resolution of the collective action problems they pose is Ostrom, *supra*.

[45] See Ostrom, *supra*.

[46] Carol M. Rose, The Several Futures of Property: Of Cyberspace and Folk Tales, Emission Trades and Ecosystems, 83 Minn. L. Rev. 129 (1998).

[47] Michael A. Heller, The Tragedy of the Anticommons: Property in the Transition from Marx to Markets, 111 Harv. L. Rev. 621 (1998). I refer to Heller, rather than to Michelman, who to the best of my knowledge coined the term, see Frank I. Michelman, Ethics, Economics, and the Law of Property, *in* Nomos XXIV: Ethics, Economics, and the Law 3 (J. Roland Pennock & John W. Chapman eds., 1982), because the concept, applied to inefficiently defined property rights relative to the efficient boundaries of resources as opposed to resources as to which everyone has a right to exclude, took off with Heller's use rather than earlier.

[48] Carol M. Rose, Left Brain, Right Brain and History in the New Law and Economics of Property, 79 Org. L. Rev. 479 (2000).

[49] Ostrom, Governing the Commons, is the most comprehensive survey. Anther seminal study was James M. Acheson, The Lobster Gangs of Maine (1988). A brief intellectual history of the study of common resource pools and common property regimes can be found in Charlotte Hess & Elinor Ostrom, Artifacts, Facilities, And Content: Information as a Common-pool Resource, (paper for the "Conference on the Public Domain," Duke Law School, Durham, North Carolina, November 9-11, 2001).

[50] The particular focus on social norms rather than formal regulation as central to the sustainability of common resource pool management solutions that are not based on property is Ellickson's, *supra*.

Peer production of information entails purely a provisioning problem. Because information is nonrival, once it is produced no allocation problem exists. Moreover, provisioning of information in a ubiquitously networked environment may present a more tractable problem than provisioning of physical matter, and shirking or free riding may not lead quite as directly to non-production.  First, the modularity and granularity of the projects suggests that occasional defections can be overcome by redundant provisioning, and will not threaten the whole.  Second, the likelihood of free riding increases as the size of the pool increases and the probability of social-norms-based elimination of free riding declines.[52]  But as the size of the pool increases, the project can tolerate increasing levels of free riding as long as the absolute number of contributors responding to individually appropriated gains—pleasure, human capital, reputation etc.—remains sufficient to pool the effort necessary to produce the good.  Indeed, for those who seek indirect appropriation like reputation, human capital, or service contracts, a high degree of use of the end product (including by "free riders" who did not contribute to writing it) increases the social value of the product, and hence the reputation, human capital, and service market value of contribution.  Third, the public goods nature of the product means that free riding does not affect the capacity of contributors to gain full use of their joint product, and does not degrade their utility from it.  This permits contributors who contribute in expectation of the use value of the good to contribute without concern for free riding.

A number of types of defection that would affect either motivation to participate or the efficacy of participation could however, affect provisioning.  The former covers actions that could reduce the intrinsic value of participation or the expected extrinsic value contributors expect to reap.  The latter relates to potential failures of integration, due to an absence of an integration process, or due to poor quality contributions, for example.  These are the kinds of defection a peer production process must deal with if it is to be successful.

There are two kinds of actions that could reduce the intrinsic benefit of participation.  First is the possibility that behavior will affect the contributors' valuation of the intrinsic value of participation.  Two primary sources of negative effect seem likely. The first is a failure of integration, so that the act of individual provisioning is seen as being wasted, rather than adding some value to the world. This assumes that contributors have a taste that places some positive value on contributing to a successful project. If this is not the case—if integration is not a component of the intrinsic value of participation—then failure to integrate would not be significant. The World Wide Web is an example where it is quite possible that the putting a web site on a topic one cares about is sufficiently intrinsically valuable to the author, even without the sense of adding to the great library of the web, that integration is irrelevant to the considerations of many contributors.

---

[51] Ostrom, Governing the Commons, 69-88.

[52] On the relationship between how small and closely knit a group is, and its capacity to use social norms to regulate behavior see Robert C. Ellickson, Order Without Law (1991).  On the importance of social norms in regulating behavior generally, and how it relates to regulation of behavior in cyberspace see Lawrence Lessig, Code and Other Laws of Cyberspace (1999).

The second, and most important potential "defection" from commons-based peer production, is unilateral[53] appropriation. Unilateral appropriation could, but need not, take the form of commercialization of the common efforts for private benefit. More directly, appropriation could be any act where an individual contributor tries to make the common project reflect his or her values too much, thereby alienating other participants from the product of their joint effort. The common storytelling enterprise called LambdaMOO, and the well-described crises that it went through with individuals who behaved in sundry antisocial ways—like forcing female characters to "have sex" that they did not want to have in the story[54]—is a form of appropriation—taking control to have the joint product serve one's own goals. In LambdaMOO the participants set up a structure for clearing common political will in response to this form of appropriation.[55] Similarly, some of the software-based constraints on moderation and commenting on Slashdot and other sites have the characteristic of preventing anyone from taking too large a role in shaping the direction of the common enterprise, in a way that would reduce the perceived benefits of participation to many others.

Another form of appropriation that could affect valuation of participation is simple commercialization for private gain. The effect is motivational, in the sense that it will create a sucker's reward aspect to participation in a way that, if the joint product remains free for all to use, and no one takes a large monetizable benefit, it would not. This effect would be consistent with (though not identical too) the "crowding-out" phenomenon, thought to associate the introduction of commodified sources of provisioning of certain goods—like blood—with a decline in their provisioning by volunteers.[56]   Clearly this effect is not particularly important in free software production, which has seen billions made by small contributors and nothing but honor made by the leaders of major projects.  But it is not implausible to imagine that individuals would be more willing to contribute their time and effort to NASA or a nonprofit enterprise than to a debugging site set up by Microsoft.  Whether this effect exists, how strong it is, and what are the characteristics of instances where it is or is not important is a valuable area for empirical research.

---

[53] As opposed to collective, as in the conversion of some aspect of the commons to a common property regime where high quality or consistent contribution to the commons could become a criterion for membership.

[54] Larry Lessig, Code and other Laws of Cyberspace (2000).

[55] See Julian Dibbell, A Rape in Cyberspace or How an Evil Clown, a Haitian Trickster Spirit, Two Wizards, and a Cast of Dozens Turned a Database Into a Society, The Village Voice, December 21, 1993, pages 36 through 42, text available ftp://ftp.lambda.moo.mud.org/pub/MOO/papers/VillageVoice.txt.

In addition to intrinsic value of participation, there is also an important component of motivation that relies on the use value of the joint project and on indirect appropriation based on continued access to the joint product—service contracts, human capital etc.  For such projects, defection again may take the form of appropriation, in this case by exclusion of the contributors from the use value of the end product.  (Why academics, for example, are willing to accept the bizarre system in which they contribute to peer review journals for free, sometimes even paying a publication fee, and then have their institutions buy this work back from the printers at exorbitant rates remains a mystery.)  In free software, the risk of defection through this kind of appropriation is deemed a central threat to the viability of the enterprise, and both the more purist licenses on the style of the GNU GPL and the more accommodating open source licenses, at least those that comply with the open source definition, prevent one person from taking from the commons, appropriating the software, and excluding others from it.  This creates a problem that, on its face, looks like an allocation problem—one person is taking more than their fair share.  But again, this is true only in a metaphoric sense.  The good is still public, and is physically available to be used by everyone.  Law (intellectual property) may create this "allocation problem," but the real problem is effect on motivation to provision, not an actual scarcity that requires better allocation. The risk of appropriation lowers the expected value contributors can capture from their own contribution, and hence lowers motivation to participate and provide the good.

Third, there is the problem of provisioning the integration function itself.  It is important to understand from the discussion here that integration requires some process for assuring the quality of individual contributions.  This could take the form of (a) hierarchically managed review, as in the Linux development process, (b) peer review, as in the process for moderating Slashdot comments, or (c) aggregation and averaging of redundant contributions.   Academic peer production of science is traditionally some combination of the former two, although the Los Alamos Archive[57] and the Varmus proposal for changing the model of publication in the health and biomedical sciences[58] towards free online publication coupled with post-publication peer commentary as a check on quality would tend to push the process further towards pure peer review.

---

[56] See Osterloh & Frey, *supra*.  Obviously the crowding out effect is different, in that there the possibility of commercialization by the contributors leads them to lose motivation, but presumably the availability of a third party who will commerialize, rather than the opportunity to commercialize oneself, will likely have greater effects of the same variety.

The first thing to see from the preceding six paragraphs is that provisioning integration by permitting the integrator to be the residual owner (in effect, to "hire" the contributors and act as the entrepreneur) presents substantial problems for the motivation to provision in a peer-based production model—both intrinsic and extrinsic motivation. Appropriation may so affect motivation to participate that the residual owner will have to resort to market- and hierarchy-based organization of the whole production effort. Second, property rights in information are always in some measure inefficient. Creating full property rights in any single actor whose contribution is only a fraction of the overall investment in the product is even less justifiable than doing so for a person who invests all of the production costs. Third, and related, integration is quite possibly, particularly with the introduction of software-based management of the communications and to some extent the integration of effort, a low-cost activity. To the extent that this is so, even though integration may require some hierarchy, or some market-based provisioning, it is a function that can nonetheless be sustained on low returns—be it by volunteers, like those who run integration of code into Linux, by government agencies, like NASA clickworkers, or by firms that appropriate the value they add through integration by means less protected from competition than intellectual property rights-based business models.

The cost of integration—and hence the extent to which it is a limit on the prevalence of peer production—can be substantially reduced by both automation and the introduction of an iterative process of peer production of integration itself. First, integration could be a relatively automated process for some products. NASA clickworkers' use of automated collation of markings and averaging out of deviations is an example, as are many of the attributes of Slashdot. Second, the integration function itself can be peer produced. Again with Slashdot, the software that provides important integration functions is itself an open-source project—in other words, peer-produced. The peer review of the peer reviewers—the moderators—is again distributed, so that 90% of registered users can review the moderators, who in turn review the contributors. As peer production is iteratively introduced to solve a greater portion of the integration function, the residual investment in integration that might require some other centralized provisioning becomes a progressively smaller investment, one capable of being carried on by volunteers or by firms that need not appropriate anything approaching the full value of the product.[59]

---

[57] Katie Hafner, Physics on the Web Is Putting Science Journals on the Line, NYT April 21, 1998 Sec. F. p. 3, col 1.

[58] Harold Varmus, E-BIOMED: A Proposal for Electronic Publications in the Biomedical Sciences (1999) http://www.nih.gov/about/director/pubmedcentral/ebiomedarch.htm

Moreover, integration, not only or even primarily integration into a general product but integration as a specific customization for specific users, could also provide an opportunity for cooperative appropriation.[60] There are no models for this yet, but the idea is that many peers will be admitted to something that is more akin to a common property regime than a commons, probably on the basis of reputation in contributing to the commons, and these groups would develop a system for receiving and disseminating service/customization projects (if it is a software project) or other information production processes. This would not necessarily work for all information production, but it could work in some. The idea is that the indirect appropriation itself would be organized on a peer model, so that reputation would lead not to being hired as an employee by a hierarchical firm, but would instead be performed by a cooperative, managed and "owned" by its participants. Just as in the case of Slashdot, some mechanism for assuring quality of work in the products would be necessary, but it would be achievable on a distributed, rather than a hierarchical model, with some tracking of individual contribution to any given project (or some other mechanism for distribution of revenues). The idea here would be to provide a peer-based model for allowing contributors to share the benefits of large-scale service projects, rather than relegating them to individual appropriation based on whatever comes down the pike.

The extent to which integration can be provided in a manner that does not require appropriation by the integrator is the third limiting factor—in addition to modularity and granularity of components of the product—on whether or not a given type of information good can be produced on a peer production model. The extent to which it is an efficient limit depends on how sensitive a project is to integration, and how well integration can be performed using technology, iterative peer production, social norms, or non-appropriating market or hierarchical mechanisms.

---

[59] Boyle focuses on this characteristic as the most interesting and potentially important solution. See Boyle, Second Enclosure Movement, *supra*.

[60] I owe the idea of cooperative appropriation to an enormously productive conversation with David Johnson. It was his idea that the peer production model can be combined with the producers' cooperative model to provide a mechanism of appropriation that would give contributors to peer production processes a more direct mechanism for keeping body and soul together while contributing, rather than simply awaiting for reputation gains to be translated into a contract with a company.

## Conclusion

In this paper I suggest that peer production of information is a phenomenon of much broader economic implications for information production than thinking of free software alone would suggest. I describe peer production common enterprises occurring throughout the value chain of information production on the Net, from content production, through relevance and accreditation, and to distribution. I also suggest that peer production has a systematic advantage over markets and firms in matching the best available human capital to the best available information inputs to create the most desired information products. The paper suggests that if peer production has a sufficient advantage in terms of its capacity to process information about who the best person is for a given information production job over firm and market based mechanisms to outweigh the costs of coordination, then peer production will outperform firms and markets.

I suggest that peer production of information is emerging because the declining price of physical capital involved in information production, and the declining price of communications lower the cost of peer production, and make human capital the primary involved economic good. This both lowers the cost of coordination and increases the importance of the factor at which peer production has a relative advantage—identifying the best available human capital for a job in very refined increments. If this is true it would have a number of implications both for firms seeking to structure a business model for the Net, and for governments seeking to capitalize on the Net to become more innovative and productive.

For academics, peer production provides a rich area for new research. Peer production, like the Net, is just emerging. While there are some studies of peer-produced software, there is little by way of systematic research into peer production processes more generally. There is much room for theoretical work on why they work, what are potential pitfalls, and what are solutions that in principle and in practice can be adopted. The role of norms, the role of technology, and the interaction between volunteerism and commercial gain in shaping the motivation and organization of peer production are also important areas of research. Qualitative and quantitative studies of the importance of peer production in the overall information economy, and in particular the Internet-based information economy would provide a better picture of just how central or peripheral a phenomenon this is.

For firms, the emergence of peer production may mean a more aggressive move from a product-based business model to a service based business model. Businesses could, following IBM and Red Hat in open source software, focus their "production" investment in providing opportunities for peer production, aiding in that production, and performing if necessary some of the integration functions. Firms that adopt this model, however, will not be able to count on appropriating the end product directly, because the threat of appropriation will largely dissipate motivations for participation. Indeed, the capacity of a firm to commit credibly *not* to appropriate the joint project will be crucial to its success in building a successful business model alongside a peer production process. This would require specific licenses that secure access to the work over time to contributors and all. It would also require a business model that depends on indirect appropriation of the benefits of the product.[61] Selling products or services, for which availability of the peer-produced product increases demand, as in the case of IBM servers that run Linux and Apache software, could do this. Conversely, firms that benefit on the supply side from access to certain types of information can capitalize on peer production processes to provide that input cheaply and efficiently, while gaining the firm-specific human capital to optimizing their product to fit the information. Again, IBM's investment in engineers who participate in writing open source software releases it from reliance on proprietary software owned by other firms, thereby creating supply side economies to its support of peer production of software. Similarly, NASA's utilization of peer production saves on its costs of mapping Mars craters. Another option is sale of the tools of peer production itself, for example, the software and access to a massive multiplayer online game like Ultima Online.

For regulators, the implications are quite significant. In particular, the current heavy focus on strengthening intellectual property rights is exactly the wrong approach to increasing growth through innovation and information production if having a robust peer production sector is important to an economy's capacity to tap its human capital efficiently. Strong intellectual property rights, in particular rights to control creative utilization of existing information, harm peer production by raising the cost of access to existing information resources as input. This limits the capacity of the hundreds or thousands of potential contributors from considering what could be done with a given input, and applying themselves to it without violating the rights of the owner of the information input. This does not mean that intellectual property rights are all bad. But we have known for decades that intellectual property entails systematic inefficiencies as a solution to the problem of private provisioning of the public good called information. The emergence of commons-based peer production adds a new source of inefficiency.

The strength of peer production is in matching human capital to information inputs to produce new information goods. Strong intellectual property rights inefficiently shrink the universe of existing information inputs that could be subjected to this process of matching human capital. Instead, owned inputs will be limited to human capital with which the owner of the input has a contractual relationship. Moreover, the entire universe of peer-produced information gains no benefit from strong intellectual property rights. Since the core of commons-based peer production is provisioning without direct appropriation, and since indirect appropriation—be it intrinsic gains from participation, or indirect extrinsic gains of reputation, human capital, serving associated demand, etc.—does not rely on control of the information, but on its widest possible availability, intellectual property offers no gain, and only loss, to peer production. While it is true that free software currently uses copyright-based licensing to prevent certain kinds of defection from peer production processes, the same protection from defection could be provided by creating a public mechanism for contributing one's work in a way that makes it unsusceptible to downstream appropriation—a conservancy of sorts. Regulators concerned with fostering innovation may well better spend their efforts on providing the institutional tools that will enable thousands of people to collaborate without appropriating the joint product, and making the public good of information they produce publicly available, than spending their efforts as they now do, increasing the scope and sophistication of the mechanisms for private appropriation of this public good.

---

[61] For a general mapping of indirect appropriation mechanisms see Yochai Benkler, Intellectual Property and the Organization of Information Production, *forthcoming* Int'l J. L. & Ec., 2002, http://www.law.nyu.edu/benklery/IP&Organization.pdf.

That we cannot fully understand a phenomenon does not mean that it does not exist. That a seemingly growing phenomenon refuses to fit our settled perceptions of how people behave and how economic growth occurs counsels closer attention, not studied indifference and ignorance. Peer production presents a fascinating phenomenon that could allow us to tap unknown reserves of human creative effort. It is of central importance to policy debates today that we not squelch it, or, more likely, move its benefits to economies that do appreciate it and create the institutional conditions needed for it to flourish.